\pdfoutput=1
\documentclass[journal=jacsat,manuscript=article,layout=twocolumn]{achemso}

\usepackage[version=3]{mhchem} % Formula subscripts using \ce{}
\usepackage[T1]{fontenc}       % Use modern font encodings
\usepackage[usenames,dvipsnames]{xcolor}
\usepackage{tabularx}
\usepackage[colorlinks=true,
	    linkcolor=red,
	    citecolor=blue,
	    filecolor=black,
	    urlcolor=black,
	    pdftex,
            pdfauthor={I. Adroher-Benitez, A. Moncho-Jorda, J. Dzubiella},
            pdftitle={Sorption and spatial distribution of protein globules in charged hydrogel particles},
            pdfsubject={Manuscript}
            ]{hyperref}

\author{Irene Adroher-Ben\'{\i}tez}

\author{Arturo Moncho-Jord\'{a}}
\affiliation[Universidad de Granada]{Departamento de F\'{\i}sica Aplicada, Facultad de Ciencias, Universidad de Granada, Avenida Fuentenueva 2, 18001, Granada, Spain}
\alsoaffiliation[Instituto Carlos I de F\'{\i}sica Te\'{o}rica y Computacional]{Instituto Carlos I de F\'{\i}sica Te\'{o}rica y Computacional, Universidad de Granada, Avenida Fuentenueva 2, 18001, Granada, Spain}

\author{Joachim Dzubiella}
\affiliation[Institut f\"{u}r Physik]{Institut f\"{u}r Physik, Humboldt-Universit\"{a}t zu Berlin, Newtonstr. 15, D-12489 Berlin, Germany}
\alsoaffiliation[Institut f\"{u}r Weiche Materie and Funktionale Materialen]{Institut f\"{u}r Weiche Materie and Funktionale Materialen, Helmholtz-Zentrum Berlin, Hahn-Meitner Platz 1, D-14109 Berlin, Germany}
\alsoaffiliation[Multifunctional Biomaterials for Medicine]{Multifunctional Biomaterials for Medicine, Helmholtz Virtual Institute, 14513 Teltow, Germany}
\email{joachim.dzubiella@helmholtz-berlin.de}
\title[Sorption and spatial distribution of protein globules in charged hydrogel particles]{Sorption and spatial distribution of protein globules in charged hydrogel particles}


\def\b{\beta}
\def\s{\sigma}
\def\k{\kappa}
\def\zi{z_{i}}

\def\lB{\lambda_{\mathrm{B}}}
\def\lD{\lambda_{\mathrm{D}}}
\def\epso{\varepsilon_{0}}
\def\epsr{\varepsilon_{\mathrm{r}}}
\def\rs{\rho_{\mathrm{s}}}
\def\cs{c_{\mathrm{s}}}
\def\csbulk{c_{\mathrm{s}}^{\mathrm{bulk}}}
\def\rh{\rho_{\mathrm{h}}}
\def\cho{c_{\mathrm{h}}^{0}}
\def\ch{c_{\mathrm{h}}}
\def\cplus{c_{+}}
\def\cminus{c_{-}}
\def\cp{c_{\mathrm{p}}}		% p = polymer
\def\Rh{R_{\mathrm{h}}}
\def\kB{k_{\mathrm{B}}}
\def\e{\mathrm{e}}
\def\d{\mathrm{d}}
\def\epsrf{\mathrm{erf}}
\def\pD{\psi_{\mathrm{D}}}
\def\Vmono{V_{\mathrm{mono}}}
\def\Vmonoo{V_{\mathrm{mono}}^{0}}
\def\Vdip{V_{\mathrm{dip}}}
\def\VBorn{V_{\mathrm{Born}}}
\def\VBorno{V_{\mathrm{Born}}^{0}}
\def\Vosm{V_{\mathrm{osm}}}
\def\Vosmo{V_{\mathrm{osm}}^{0}}
\def\Velec{V_{\mathrm{elec}}}
\def\Vspec{V_{\mathrm{spec}}}
\def\Vspeco{V_{\mathrm{spec}}^{0}}
\def\Bb{\mathcal{B}_{2}}
\def\Vtot{V_{\mathrm{total}}}
\def\yo{y_{0}}
\def\qb{q_{\mathrm{b}}}
\def\zb{z_{\mathrm{b}}}
\def\mub{\mu_{\mathrm{b}}}
\def\mubb{\widetilde{\mu}_{\mathrm{b}}}
\def\Rb{R_{\mathrm{b}}}
\def\Vb{V_{\mathrm{b}}}

\begin{document}

\begin{abstract}
We have theoretically studied the uptake of a non-uniformly charged biomolecule, suitable to represent a globular protein or a drug, by a charged hydrogel carrier in the presence of a 1:1 electrolyte. Based on the analysis of a physical interaction Hamiltonian including monopolar, dipolar and Born (self-energy) contributions derived from linear electrostatic theory of the unperturbed homogeneous hydrogel, we have identified five different sorption states of the system, from complete repulsion of the molecule to its full sorption deep inside the hydrogel, passing through meta- and stable surface adsorption states. The results are summarized in state diagrams that also explore the effects of varying the electrolyte concentration, the sign of the net electric charge of the biomolecule, and the role of including excluded-volume (steric) or hydrophobic biomolecule-hydrogel interactions. We show that the dipole moment of the biomolecule is a key parameter controlling the spatial distribution of the 
globules. In particular, biomolecules with a large dipole moment tend to be adsorbed at the external surface of the hydrogel, even if like-charged, whereas uniformly charged biomolecules tend to partition towards the internal core of an oppositely-charged hydrogel. Hydrophobic attraction shifts the states  towards internal sorption of the biomolecule, whereas steric repulsion promotes surface adsorption for oppositely-charged biomolecules, or the total exclusion for likely-charged ones. Our results establish a guidance for the spatial partitioning of proteins and drugs in hydrogel carriers, tuneable by hydrogel charge, pH and salt concentration.
\end{abstract}


%-------------------------------------------------------------------------------------------------%
% Introduction
%-------------------------------------------------------------------------------------------------%
\section{Introduction}

Hydrogels are soft colloidal particles of nanometric size formed by cross-linked polymer chains dispersed in water. 
They have received considerable attention during the last decades due to their exceptional physicochemical properties~\cite{Stuart2010}. Firstly, hydrogels can be considered as multi-responsive nanomaterials because they are able to reversible swell and shrink in a useful and reproducible manner in response to various stimuli from their surroundings, such as variations of the the solvent quality, temperature, salt concentration, pH, or external electric/magnetic fields\cite{Murray1995,Saunders2009,Fernandez-Nieves2011}. In the swollen state, hydrogels are hydrophilic and incorporate a large amount of water, leading to very open porous structures that allow the permeation of ions and other kind of cosolutes, such as proteins, peptides, lipids, enzymes, genetic material, drugs and chemical reactants. Conversely, in the shrunken state, the particles partially expel their content due to the collapse of the cross-linked network induced by the enhanced hydrophobic attractions between the polymers. The 
release of the entrapped molecule can be triggered externally in a controlled fashion~\cite{Malmsten2010,Ramos2011,Ramos2014,Lesher-Perez2016}. Moreover, hydrogels can be designed to be biocompatible, biodegradable and, due to the large water content allow the incorporation of biomacromolecules with relatively small changes in the native structure. This preserves the drug's biological activity and conformation state, reducing the toxicity and enhancing its protection from chemical and enzymatic degradation~\cite{Frokjaer2005,Malmsten2006,Ghugare2009,Vinogradov2010,Bae2013}. Due to the combination of all these features, hydrogels have been proposed as excellent candidates for transport and delivery systems of biomacromolecules and drugs~\cite{Kim1992}, e.g.,  in anti-cancer and gene therapy, permitting a high payload capacity.

The uptake of biomolecules is mediated by the underlying physical interactions between the biomolecule and the polymer network~\cite{Yigit2012,Welsch2013}. These interactions not only determine the net degree of uptake but also the preferential location of the biomolecule in the hydrogel volume. In this respect, it must be emphasized that the properties of the hydrogel-biomolecule complex strongly depend on whether the molecule is superficially adsorbed at the external shell of the hydrogel or internally absorbed deep inside the polymer network. Protein surface adsorption in fact leads to a protein \textquoteleft corona' that largely defines the biological identity of the particle~\cite{Cedervall2007}. In some practical situations, surface adsorption is unwanted. An example would be the surface deposition of lysozyme proteins on contact lenses, which can cause adverse responses and shortens the time that  lenses can be worn~\cite{Garrett1998}. Analogously, the use of hydrogels as nanocarriers requires in 
some  cases the complete encapsulation of the therapeutic agent in the internal matrix of the particle. In this way the biomolecule is not able to interact with the biological environment, thus avoiding the enzymatic breakdown before reaching the site of action in the body~\cite{Johansson2010}. In some other circumstances surface adsorption is desirable. For instance, the exposure of certain protein domains located at the hydrogel surface (corona)  can be used to activate specific recognition pathways for cellular uptake~\cite{Giudice2016,Shaw2016,Schottler2016}. Clearly, the details of local interactions and spatial partitioning of the drugs in hydrogels will also affect their time-dependent uptake\cite{Angioletti-Uberti2014} and release kinetics~\cite{Kim1992}.

Therefore, it is of fundamental importance to know the hydrogel-biomolecule interactions implied in the uptake process in order to understand how drugs partition in the hydrogel carrier. In general, this interaction depends on many parameters that involve solvent properties (such as temperature, electrolyte concentration, pH), hydrogel features (e.g., charge distribution, network morphology, hydrophobicity) and properties of the biomacromolecule (e.g., size, shape and charge distribution). In particular, the electrostatic interaction is shown to be one of the most relevant contributions in biomolecule uptake, as, e.g., shown in experiments performed with different types of peptides in the presence of oppositely-charged poly(acrylic acid) or poly(acrylamidopropyltriethylammo-niumchloride) hydrogels~\cite{Bysell2009,Bysell2009b,Hansson2012}. Here, the incorporation is enhanced simply by increasing the peptide charge. 
Similarly, the encapsulation of Cytochrome C proteins inside oppositely-charged hydrogels is also electrostatically driven, leading to a uniform distribution of proteins within the structure~\cite{Smith2011}. Naturally, salt concentration and pH become then additional factors affecting the uptake process. On the one hand, lowering the electrolyte concentration enhances both the electrostatic interactions and the osmotic repulsion induced by the \textit{free} counterions confined within the hydrogel~\cite{Yigit2012,Yigit2017}.
On the other hand, the presence of pH-sensitive functional groups allows to control the sign (as well as the distribution) of the charge in the biomacromolecule. In this work we show that the charge tuning due to pH-sensitivity may play a determinant role in the total hydrogel-biomolecule interaction.
For instance, close to  the isoelectric point of the molecule, charge regulation can even lead to charge inversion of the biomolecule~\cite{Biesheuvel2005, Lund2005}.

Additional experimental studies evidence that charged proteins can significantly partition into planar and spherical polyelectrolyte brushes even when they hold the same net charge, at pH conditions far away from the isoelectric point~\cite{Wittemann2003,Wittemann2006,Ballauff2006,Henzler2010}. This adsorption \textquoteleft on the wrong side' is surprising in the sense that, intuitively, proteins are expected to be repelled by electrostatic and excluded-volume forces. Recent theoretical studies and simulations have shown that this phenomenon can be attributed to the superposition of several interactions.
First, proteins and other biomacromolecules usually carry patches of opposite charge sign.\cite{Leermakers2007} The non-uniform, \textquoteleft multipolar' charge distribution generates an effective dipole moment that interacts asymmetrically with the polyelectrolyte, leading to an attraction that can overcome the electrostatic repulsion~\cite{Hu2009,Yigit2017}. Second, due to the larger electrolyte concentration found inside the brush compared to the bulk concentration, there is an attractive Born (self-)electrostatic energy that arises upon insertion into the charged brush~\cite{Yigit2012,Yigit2017}.
Finally, there can be additional attraction related to the release of the condensed counterions in the case of highly charged polyelectrolyte chains: when the protein patches of opposite charge come into contact with the chains, the release of the counterions leads to a large increase of the translational entropy in the system~\cite{Welsch2013,Wittemann2006,Yigit2017,Leermakers2007}. All these effects are also expected to be present at some extent in the sorption of biomacromolecules to charged hydrogels. 

In addition to electrostatic forces, the spatial partitioning of the biomolecule inside the hydrogel will be influenced by other sorts of interactions, such as steric repulsion effects or hydrophobic attractions. Excluded-volume effects, for instance, have been observed in the encapsulation of proteins and peptides when the mean network mesh size is smaller than the size of the biomolecule~\cite{Garrett1998,Bysell2006}. Hence, the biomolecules are precluded to enter the hydrogel and may tend to adsorb at the surface as the molecular weight of the biomolecule is raised, the cross-linker concentration within the hydrogel increases, or the hydrogel de-swells. Otherwise, the hydrophobic attraction between the incoming biomolecule and the hydrogel can be a very important driving force for some specific macromolecules or drugs. Indeed, increasing both the biomolecule and hydrogel hydrophobicity significantly enhances the sorption to the hydrogel, where the preferential location of the binding depends on the 
swelling state of the nanoparticle~\cite{Welsch2013,Kawaguchi1992,Lord2006,Lord2006b,Bysell2010,Bysell2010b}.

The aim of this work is to explore theoretically the conditions that guarantee stable and metastable uptake of biomolecules and, in particular, how the latter spatially partition within the hydrogel (e.g., surface adsorption versus sorption deep inside). Our analysis is based on a well-defined interaction Hamiltonian~\cite{Yigit2012, Angioletti-Uberti2014} extended to include dipolar contributions~\cite{Yigit2017} between a non-uniformly charged biomolecule and the charged hydrogel particle.
This allows to consider many different types of biomolecules attending to their net charge and dipole moment (charge asymmetry), both controllable by pH. The potential of mean force exerted by the hydrogel as a function to the distance (i.e., the effective interaction Hamiltonian) is determined for different combinations of net charge, dipole moment, bulk salt concentration, and hydrophobic/steric effects. The calculations include the effect of the local variation of the electrolyte concentration from the  bulk reservoir to the internal volume of the hydrogel, which has important implications especially for low salt concentrations. We finally summarize in state diagrams where the biomacromolecule uptake is more likely to occur, i.e., classify the results in five different states, namely repulsion (no adsorption), metastable surface adsorption, stable surface adsorption, and partial and full inside sorption. 

\section{Model and method}

\subsection{Model of the hydrogel and globular biomolecule}

The system under study consists of a charged hydrogel particle and a single charged dipolar biomolecule in an aqueous suspension in presence of 1:1 electrolyte at a given bulk concentration $\csbulk$. The biomolecule is modeled by a spherical particle of radius $\Rb$, representing for instance a protein globule. Analogously, the hydrogel is represented by a penetrable sphere of radius $\Rh$, and assumed to be in the swollen state.
The inner structure of the polymer network of the hydrogel may present many different configurations. %\cite{Daly2000,Lopez-Leon2006,Geisel2015}.
For instance, a wide range of hydrogel particles with a highly cross-linked core and a lightly cross-linked shell\cite{Daly2000,Lopez-Leon2006},
and even more exotic configurations such as hollow microgels\cite{Geisel2015} have been reported in literature.
The ionic groups of the polyelectrolyte network may also distribute differently, either homogeneously within the hydrogel or forming an external charged shell. Nevertheless, as a start we have assumed the most general situation of a core-shell morphology, and considered that both neutral monomers and charged groups follow a uniform (i.e., fully homogeneous continuum) distribution inside the hydrogel particle (core). The charged and neutral monomer density smoothly decreases to zero at the hydrogel interface (shell).

\begin{figure}[ht!]
 \centering
 \includegraphics[width=3.25in]{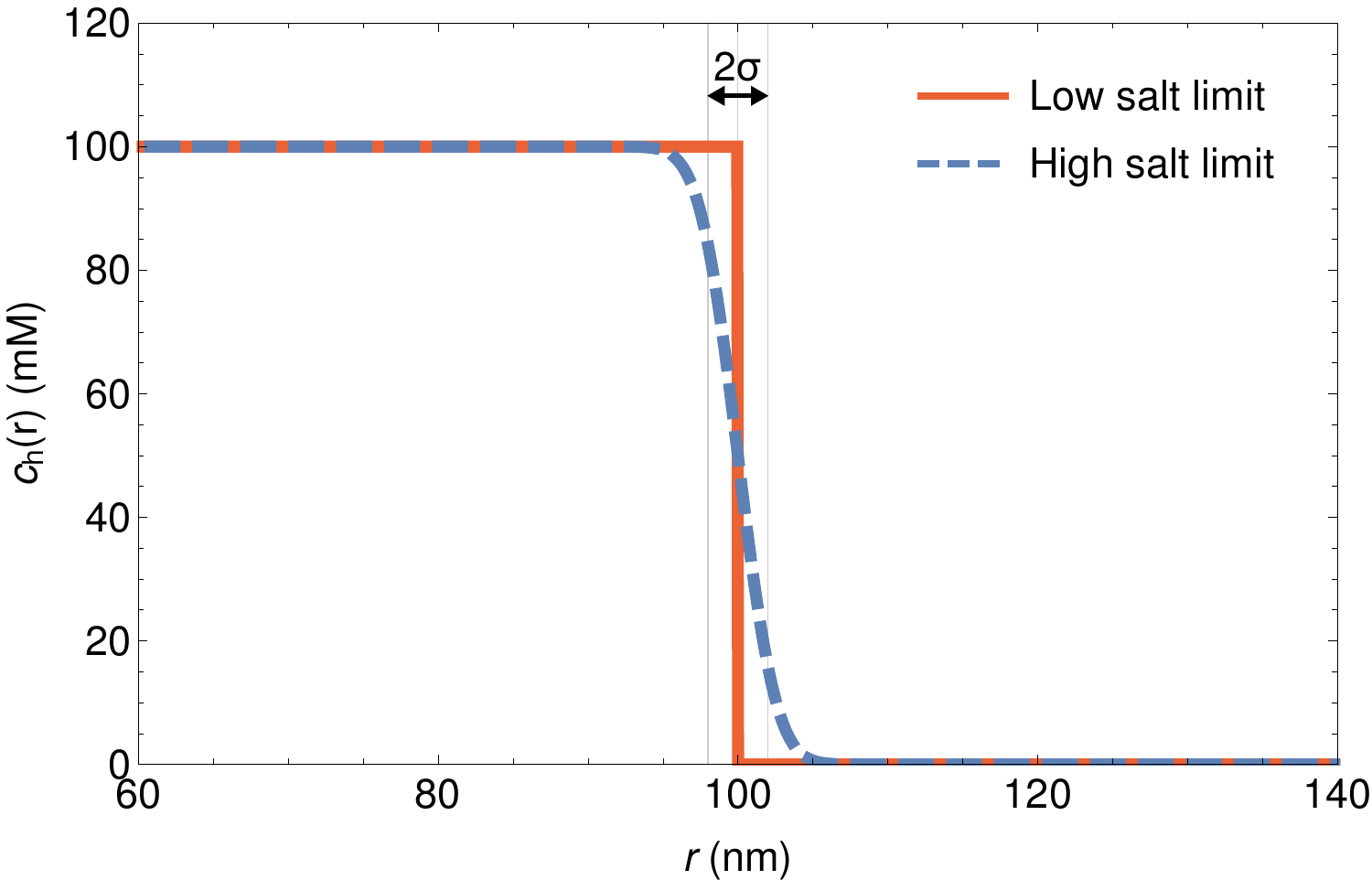}
 \caption{Model for the hydrogel charge distribution in the low and high electrolyte concentration regimes (in both cases $\cho=100$ mM and $\Rh=100$ nm).}
 \label{fig:cmrAll}
\end{figure}

Figure~\ref{fig:cmrAll} shows the hydrogel charge distribution for the two concentrations of monovalent salt studied in this work. Two limiting cases of electrolyte concentration have been considered, namely \textquoteleft low' and \textquoteleft high' salt concentration. These regimes are distinguished by the Debye screening length, $\lD$, and the intrinsic hydrogel-water interface width, $\s$. In the low salt limit, $\s \ll \lD$, the interface total charge distribution is governed by the salt, so the hydrogel shell can be modeled by a sharp edge. On the other hand, in the high salt regime the salt distribution follows closely the smoothly decaying shell density. The high salt regime ($\csbulk > 100$~mM) is of special relevance since it fairly represents the conditions found in physiological applications. For this we approximate the hydrogel charge distribution as a error function, so that
\begin{equation}
 \resizebox{0.85\linewidth}{!}{$\ch(r) = \displaystyle\frac{\cho}{2} \left[ 1 - \epsrf\left( \frac{r-\Rh}{\sqrt{2} \ \s} \right) \right], \ 0 < r < +\infty.$}
 \label{eq:cmrHighSalt}
\end{equation}
where $\cho$ is the hydrogel charge density in the central core of the particle, and $r$ is the distance to the hydrogel center. For both cases (low and high electrolyte concentration) the linearized Poisson-Boltzmann differential equation has been analytically solved to determine the density profiles of counterions and coions inside and around the hydrogel particle (see Appendix). 
The ionic concentrations determine the electrostatic and osmotic effective pair interactions between a single charged biomolecule and the hydrogel, as will be shown below.

It is important to notice that charged groups in proteins, peptides and other biomolecules are not usually located in the particle centre or homogeneously distributed either. On the contrary, these discrete charges are often asymmetrically distributed on and inside the molecules, leading to multipolar contributions to the electrostatic potential. 
To include the multipolar effect, in this work we have modeled the biomolecule as a particle with net charge $\qb$ and an electric dipole moment $\mub$ that accounts for the heterogeneous charge distribution in leading order. We also assume that the biomolecules are at infinite dilution, that is, we neglect the collective effects that arise from the interaction between biomolecules. Therefore, this study focuses on the sorption and distribution of a single globule.

\subsection{Effective interaction Hamiltonian}

To calculate the potential of mean force (PMF), i.e., the effective interaction Hamiltonian between the hydrogel and the charged biomolecule, we split the pair interaction in three different phenomenological contributions~\cite{Yigit2012,Angioletti-Uberti2014,Yigit2016}, 
\begin{equation}
 \Vtot(r) = \Velec(r) + \Vosm(r) + \Vspec(r). 
%\BVnonelec(r).
 \label{eq:BVtotSplit}
\end{equation}
The electrostatic interaction is in turn split into three additive terms, $\Velec(r) = \Vmono(r) + \Vdip(r) + \Delta \VBorn(r)$.
The first one, $\Vmono(r)$, represents the classical electrostatic monopole attraction or repulsion induced by the net charge of both the hydrogel and the biomolecule.
Since we are considering the general situation of a non-uniform distribution of charged groups within the biomolecule, additional multipolar contributions should be also taken into account.
In this work, we considered the dipolar interaction, $\Vdip(r)$, which represents the first leading term of the multipolar expansion beyond the monopolar contribution.
Explicitly, the classical electrostatic interaction up to the dipolar term is given by~\cite{Hill1986}
\begin{equation}
 \resizebox{0.89\linewidth}{!}{$\displaystyle \Vmono(r) + \Vdip(r) = \zb y(r) -\kB T \ln\left\{\frac{\sinh[\mubb Y(r)]}{\mubb Y(r)}\right\}$}
 \label{eq:BVmonoBVdip}
\end{equation}
where $\kB$ is Boltzmann constant, $T$ is the absolute temperature of the system, $\zb \equiv \qb / e$ is the biomolecule valency, $y(r)\equiv  e \psi(r)/\kB T$ is the normalized electrostatic potential calculated in the Appendix~\ref{sec:Appendix1} by means of the linear Poisson-Boltzmann approximation, $\mubb \equiv \mub/e$ is the electric dipole moment of the solute and $Y(r) \equiv e E(r)/\kB T$ is the normalized electric field generated by the charged hydrogel. 
The electrostatic dipolar term is always attractive because the heterogeneously-charged biomolecule tends to align its dipole with the electrostatic field generated by the charged hydrogel.
The third contribution, $\Delta \VBorn(r)$, represents the Born   interaction.
It characterizes the change of the self-energy difference of charging the biomolecule inside the charged hydrogel versus bulk solvent.
It is defined as the difference between the solvation energy at $r$ and the solvation energy in the bulk, $\Delta\VBorn(r) = \VBorn(r) - \VBorn(r\to\infty)$, being~\cite{McQuarrie1976,Yigit2017}
\begin{flalign}
\begin{split}
 & \VBorn(r) = \displaystyle \kB T \left\{  \frac{\lB \zb^2}{2 \Rb [\Rb \k(r)+1]}\right. + \\
 & \resizebox{0.8\linewidth}{!}{$\left.\frac{3 \lB \mubb^2}{2 \Rb^3} \frac{[\Rb \k(r)+1] [(\Rb \k(r))^2 + 2 \Rb \k(r)+2]}{[(\Rb \k(r))^2 + 3 \Rb \k(r)+3]^2} \right\}$}
 \label{eq:BVBorn}
\end{split}
\end{flalign}
where $\Rb$ is the radius of the biomolecule, $\lB \equiv e^{2}/4 \pi \epso \epsr \kB T$ is the Bjerrum length and $\k(r) = \sqrt{4\pi\lB z^{2} [c_{+}(r)+c_{-}(r)]}$ is the local inverse screening length, which depends on the local salt concentration~\cite{Yigit2012}.
The first term of equation \eqref{eq:BVBorn} is the classical monopolar result of a sphere in an electrolyte suspension in the Debye-H\"{u}ckel approximation~\cite{McQuarrie1976}.
The second term is the dipolar expansion for a dipole moment $\mubb$ centered in a sphere of radius $\Rb$.
Charging the particle inside the hydrogel is always an energy-favorable process due to the larger ionic concentration there, so the difference in Born solvation energy contributes to an attractive hydrogel-biomolecule interaction.

The second term of the right-hand side of equation \eqref{eq:BVtotSplit} stands for the effects of the osmotic pressure due to the confined ions inside the hydrogel, 
\begin{equation}
 \Vosm(r) = \kB T[\cplus(r) + \cminus(r) - 2 \csbulk] \Vb,
 \label{eq:BVosm}
\end{equation}
which depends on the biomolecule volume $\Vb$ and on the difference between local number density of free ions ($\cplus+ \cminus$) inside the polymer network and within the bulk ($2\csbulk$).
The osmotic pressure always exerts a repulsion on the incoming solute and decreases with electrolyte bulk concentration, given that the biomolecule sorption is hindered by the excess of counterions.

The third term of equation \eqref{eq:BVtotSplit} we have defined a phenomenological specific potential that accounts for the steric effects and the hydrophobic character of the biomolecule.
On the one hand, the partitioning of particles within the hydrogel is always obstructed due to the volume-exclusion exerted by the polyelectrolyte chains of the network.
This effect is naturally small for swollen hydrogels and small globules (i.e., nanometer size) but becomes very important when shrinking the hydrogel or increasing the size of the biomolecule.
However, this repulsion may be overcome by the attractive interaction that results from the hydrophobic character of the molecule. Indeed, many of the molecules employed in biotechnological applications are significantly hydrophobic and show preferential binding for the polyelectrolyte network rather than the aqueous environment.
Both steric and hydrophobic effects are considered in the following mean-field potential  
\begin{equation}
 \Vspec(r) = - \kB T \ln[1- 2\cp(r) \Bb],
 \label{eq:BVspec}
\end{equation}
where $\cp(r)$ is the monomer density of the network and $\Bb$ is the second virial coefficient (or \textquoteleft interaction' parameter) of the local two-body interaction between a biomolecule and a monomer. This potential can be derived by expanding the corresponding partitioning coefficient (for a charge neutral hydrogel)  $K=\exp(-V_{\rm spec}/\kB T)\simeq 1+\cp\Gamma$ in powers of the polymer density. The parameter $\Gamma=-2\Bb$ represents the protein-polymer adsorption in the limit of infinite dilution of solute~\cite{Shulgin2008}. If the hydrophobicity of the particle is more relevant than the excluded-volume effects, then $\Vspec(r)<0$ and the specific potential will be attractive. Otherwise, if the system exhibits strong excluded-volume effects, the overall specific interaction will be repulsive. The particular form (6) has the nice feature that it diverges for high steric constraints (i.e., close polymer packing) and penetration of the hydrogel core by the solute is strongly penalized.

An additional contribution to the biomolecule-hydrogel effective pair interaction would be the attraction induced by counterion release effects that arise due to gain of entropy associated to the release of confined counterions from the polymer network to the bulk suspension~\cite{Wittemann2006}. However, we neglect this term since it only becomes relevant for the case of strongly charged polyelectrolyte networks.~\cite{Yigit2017}

With the purpose of performing an exhaustive discussion of the forthcoming results, it is convenient to recall the main features of the energetic terms that contribute to the total PMF, and their dependence on $\qb$ and $\mub$.
First, the electrostatic monopolar contribution, $\Vmono(r)$, is attractive if the hydrogel and the biomolecule are oppositely-charged, repulsive otherwise, and does not depend on the electric dipole moment.
Since this contribution is proportional to the electrostatic potential, it reaches its maximum (absolute) value inside the hydrogel and quickly decreases in the bulk. The electrostatic dipolar term, $\Vdip(r)$, is always attractive and it is coupled to the local electric field. Hence, it becomes especially relevant at the hydrogel surface, where the electrostatic field reaches its maximum value. Since $\Vdip(r)$ is the result of the inhomogeneous charge distribution of the biomolecule, it is independent of $\qb$ and increases with $\mub$. Considering the monopolar and dipolar part of Born solvation energy together, their contribution is always attractive and increases with $\qb$ and $\mub$.
On the other hand, the osmotic pressure always exerts a repulsive interaction and does not depend on $\qb$ and $\mub$, but increases as the salt concentration drops. Finally, the specific contribution, $ \Vspec(r)$,  may be attractive, null or repulsive, as it is the result of the interplay between steric exclusion and hydrophobic effects.

%-------------------------------------------------------------------------------------------------%
% Results and discussion
%-------------------------------------------------------------------------------------------------%
\section{Results and discussion}

\subsection{PMF features and the definition of sorption states}

\begin{figure}[ht!]
 \centering
 \includegraphics[width=3.25in]{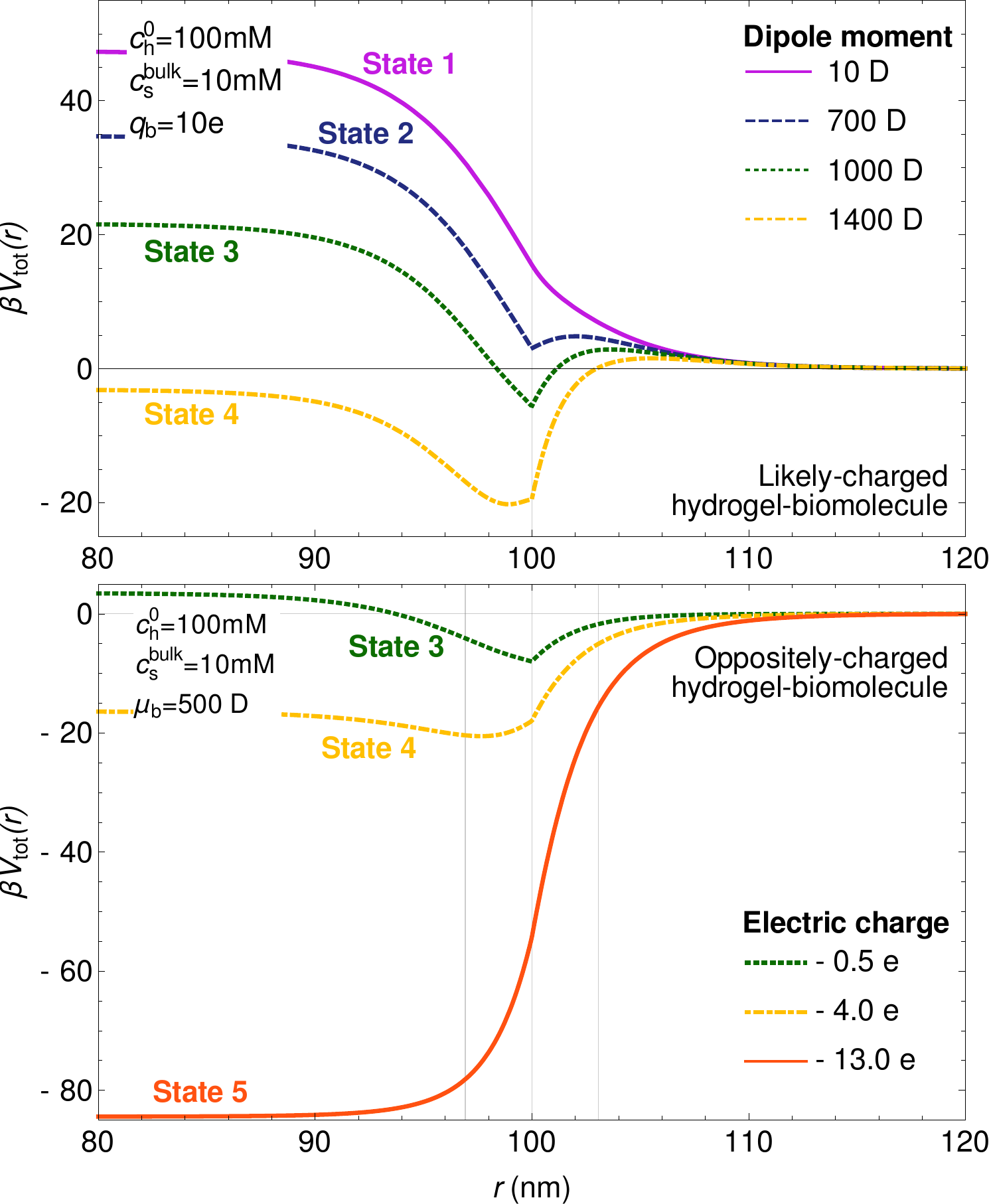}
 \caption{Potential of mean force (PMF) as a function of the distance to the hydrogelcenter, $r$, evaluated in the low salt regime ($\csbulk = 10$~mM). Top graph shows results obtained for a likely-charged biomolecule ($\qb=10e$) and varying the electric dipole moment of the biomolecule, $\mub$ (in Debye units). Bottom graph depicts the curves calculated for a oppositely-charged biomolecule, fixing the electric dipole moment to $\mub=500$~D and varying its electric charge $\qb$.
All the subsequent results are classified in terms of the five states illustrated in both graphs.}
 \label{fig:BVtot}
\end{figure}

From the shape of the PMF curves is possible to determine the sorption degree and predict where the biomolecule will be preferentially partition. In this work we have distinguished five different sorption states: (1) no adsorption, (2) metastable surface adsorption, (3) stable surface adsorption, and (4) partial and (5) full inside sorption. As an example, in Figure~\ref{fig:BVtot} a representative PMF is plotted for each state of the system. 

\begin{itemize}
\item {\bf State 1} corresponds to those cases in which the total PMF is completely repulsive, avoiding any permeation of the biomolecule within the polyelectrolyte network of the hydrogel. 

\item In {\bf state 2} a local minimum in the repulsive potential located at the hydrogel interface arises. 
This potential well allows an accumulation of biomolecules, leading to a metastable adsorption onto the hydrogel surface. 

\item In {\bf state 3}  the PMF is repulsive inside the hydrogel, but the local minimum is deep enough to become attractive, the adsorption of the biomolecule in the hydrogel surface is stable. 

\item In {\bf state 4} the hydrogel-biomolecule interaction potential is also attractive inside the microgel, partitioning of the globules will also significantly happen inside the hydrogel network. 

\item Finally, in {\bf state 5} no surface adsorption takes place but full inside sorption. 

\end{itemize}

Many experimental evidences of these five states for related systems may be found in the literature. Representative examples of state 1 are often achieved for the case of charged proteins electrostatically repelled to likely-charged polyelectrolyte brushes or hydrogels at high enough salt concentration~\cite{Wittemann2006}. Surface adsorption (states 2 and 3) are found, for instance, in the interaction between cross-linked poly(acrylic acid) hydrogels and oppositely-charged poly-L-lysine peptides of large molecular weight~\cite{Bysell2006}. Indeed, peptides larger than the effective network pore size are unable to penetrate inside the hydrogel, and become concentrated at its surface. In addition, surface adsorption states may be also enhanced by dipolar interactions between proteins and charged reverse micelles~\cite{Pitre1993}. Partitioning between surface adsorption and diffusive penetration (state 4) has been identified in isothermal titration calorimetry of sorption of $\beta$-Lactoglobulin 
on spherical polyelectrolyte brushes at low salt concentration~\cite{Henzler2010}. In these experiments, the two sorption steps observed in the integrated heat clearly indicates the existence of two different binding sites, one internal and another in the outer region of the brush. Finally, complete inside adsorption (state 5) is generally attained in the presence of oppositely-charged proteins with small molecular size at low electrolyte concentration. Under these conditions, the strong electrostatic or hydrophobic attractions with charged polymer network lead to the total encapsulation of the biomolecule within the internal core of the particle~\cite{Garrett1998,Wittemann2006,Bysell2006}.

\subsection{Parameter range}

The sorption states of the system for a large interval of (positive and negative) net charge and electric dipole moment have been studied. It should be emphasized that pH and salt concentration may have a very important effect on the spatial partitioning of the biomolecule. On the one hand, it may modify the net charge of the biomolecule, $\qb$, and even invert its sign when passing through the isoelectric point. On the other hand, changes on pH also alter the spatial distribution of charges within the biomolecule, which in turn may induce changes in the value of the biomolecule dipole moment, $\mub$. In our work, we do not explicitly study the effect of the pH, as it would imply to assume a specific biomolecule in a particular salty condition. Instead of doing a particular choice, we try to keep the study as general as possible, and investigate the equilibrium partitioning of any kind of biomolecule in terms of $\qb$ and $\mub$. The application of our theoretical predictions to some 
experimental data will require from the experimentalist the previous determination of $\qb$ and $\mub$ of the biomolecule for the pH and salt concentration conditions of the experimental setup. From the early experimental works~\cite{Oncley1942}, to more recent computational tools working on data of the Protein Data Bank\cite{Felder2007,Berman2000}, we have determined the range of possible values of ($\mub,\qb$). We have ensured that the broadness of the intervals is enough to cover all the possible cases that may take place in most of the experimental situations. The biomolecule net charge can reach values up to $\pm20e$ (elementary charge units) because the solution pH can highly tune the electric charge, especially when the biomolecule bears pH-sensitive functional groups. Analogously, the biomolecule dipole moment varies from negligible values to $2000$~D (Debye units). Every sorption state resulting from each combination of net charge and dipole moment has been represented in ($\mub,\qb$) state 
diagrams.

As mentioned above, the electrolyte concentration has also a determinant effect on the hydrogel-molecule interaction.
Therefore, two limiting cases of low and high 1:1 electrolyte concentration have been considered. In the low salt regime we studied the system at $\csbulk = 1$ mM and $\csbulk = 10$ mM, but we observed that in the former case the osmotic pressure was so high that no sorption state was achieved in any case. Therefore, in this work we have focused at $\csbulk = 10$ mM for the low salt regime, while in the high salt limit we have considered $\csbulk = 100$ mM (note that this case is the relevant one for physiological applications). We also explored the role of the charge sign of the biomolecule. Therefore, the combination of low and high salt regimes and of oppositely- and likely-charged biomolecules leads to four different state diagrams. 

In addition, the hydrophobicity of the biomolecules and the volume exclusion exerted by the hydrogel (entering through $\Vspec(r)$) can also determine the uptake into the polyelectrolyte network. In order to investigate both effects, three different hypothetical frameworks have been analyzed: a system where both the hydrogel exclusion-volume and the biomolecule hydrophobicity are negligible ($\Vspec(r)=0$), a system in which the biomolecule hydrophobicity dominates the specific interaction ($\Vspec(r)<0$), and a system with high excluded-volume effects ($\Vspec(r)>0$). As a result, four state diagrams have been plotted for each one of the three systems in Figures~\ref{fig:PhaseDiagramBVelec0}, \ref{fig:PhaseDiagramBVelec-5} and \ref{fig:PhaseDiagramBVelec5}, respectively.
A summary of all the other system features is listed in Table~\ref{tab:data}.
\begin{table}[ht!]
 \centering
 \small
 \begin{tabularx}{\linewidth}{Xcc}
  \textbf{Variable} & \textbf{Symbol} & \textbf{Value}\\
  \hline\hline
  Absolute temperature & $T$ & 298 K \\
  Hydrogel radius & $\Rh$ & 100 nm \\
  Concentration of hydrogel charged groups at $r=0$& $\cho$ & 100 mM \\%($\yo=2.12$)\\
  Width of the hydrogel shell (high salt limit) & $\s$ & 2 nm \\
  Valence of ions & $z_\pm$ & $\pm$ 1\\
  Radius of the biomolecule & $\Rb$ & 1.5 nm\\
  \hline
  \end{tabularx}
  \caption{Summary of the values of the variables in the hydrogel-electrolyte-biomolecule system.}
  \label{tab:data}
\end{table}

\subsection{Discussion of state diagrams}

\begin{figure}[ht!]
 \centering
 \includegraphics[width=3.25in]{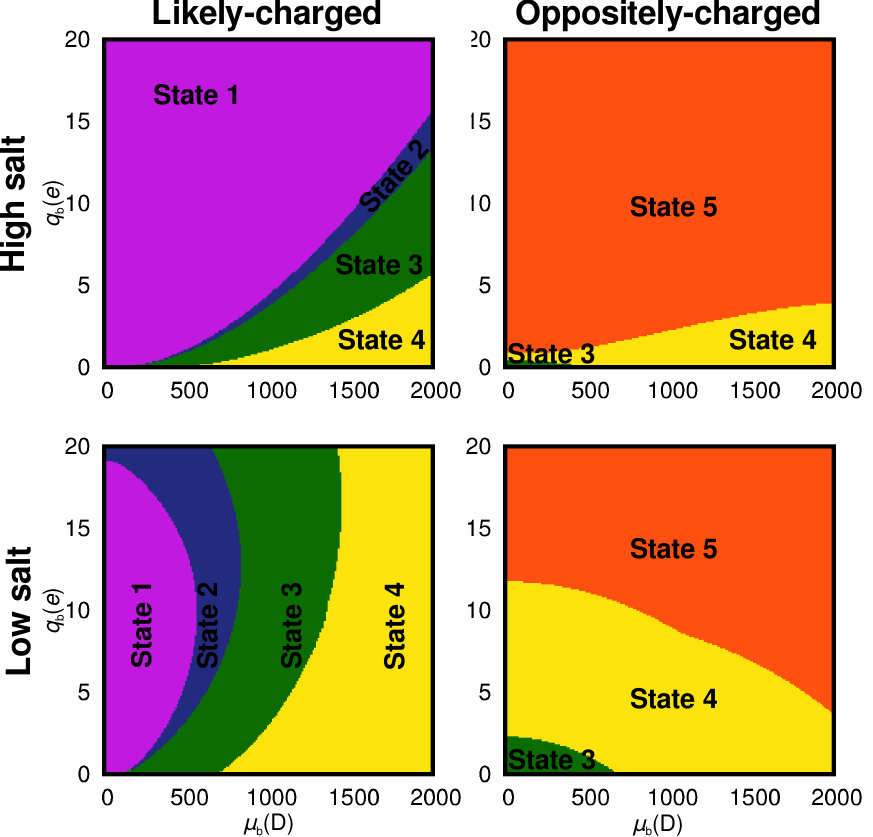}
 \caption{($\mub, \qb$) state diagrams obtained in the high and low salt concentration regimes (up and down panels, respectively) and for like-charged (left panels) and oppositely-charged (right panels) hydrogel and biomolecule in a system where steric repulsion is compensated with hydrophobic attraction ($\Vspec(r)=0$).}
 \label{fig:PhaseDiagramBVelec0}
\end{figure}

To begin with, we discuss the general trends of the state diagrams for $\Vspec(r)=0$, that is, a system where the hydrogel exclusion-volume is negligible and the biomolecule does not possess any hydrophobic character (see Figure~\ref{fig:PhaseDiagramBVelec0}). We first analyze the theoretical predictions for biomolecules oppositely-charged to the hydrogel. In such scenario, the stable sorption states 4 and 5 are resulting from almost any combination of ($\mub, \qb$), with the exception of very small values of charge and dipole moment. This is a predictable result, since all contributions to the total PMF but the osmotic interaction are attractive. Therefore, the repulsion exerted by the osmotic pressure is only relevant when $\mub$ and $\qb$ are small enough to make the other interactions less dominant. For the lower salt concentration the osmotic pressure is higher because of the larger difference between the neutralizing counterions inside the hydrogel and the counterions in the bulk.

On the other hand, in the case of likely-charged particles, the biomolecule sorption states are more dependent on the electrolyte concentration. The osmotic pressure decreases as the ionic bulk concentration increases, so it has a negligible effect on the total PMF at $\csbulk = 100$ mM. In addition, at this salt concentration the difference in the Born energy is also irrelevant for small values of ($\mub, \qb$), although it increases with both variables. Consequently, the sorption state of the biomolecule at high electrolyte concentration is mainly the result of the competition between the electrostatic monopolar repulsion (directly proportional to $\qb$) and the dipolar attraction (which increases with $\mub$). Hence, if the biomolecule net charge is high enough, monopolar repulsion will hinder its permeation deeply into the carrier independently of the electric dipole moment. However, as $\qb$ decreases and $\mub$ increases, dipolar attraction becomes more relevant, raising the chance of (meta- and stable)
 interfacial adsorption of the biomolecule, and so leading to sorption states 2 and 3. In this way, the sorption of the molecule inside the hydrogel (state 4) will be possible when $\mub$ is high enough to enhance the attraction produced by the difference in Born solvation energy.

Regarding the low electrolyte concentration regime for likely-charged particles, the sorption state diagram shows less dependence on the electrostatic monopolar repulsion than in the high salt regime. The main reason of this change is that the relevance of both the osmotic pressure and the Born solvation energy to the total PMF becomes enhanced at low electrolyte concentrations. For instance, at $\mub=1000$ D and $\qb=10e$, the osmotic contribution at the center of the hydrogel is $\Vosmo \sim 0.25 \Vmonoo$ and the Born contribution is $\Delta\VBorno \sim -0.82 \Vmonoo$ for a salt concentration $\csbulk = 10$~mM, while in the high salt regime, both contributions are only $\Vosmo \sim 0.04 \Vmonoo$ and $\Delta\VBorno \sim -0.27 \Vmonoo$ at $\csbulk = 100$~mM. Consequently, although the increase in osmotic pressure enhances the repulsion already dominated by the electrostatic monopolar interaction, the sorption state of the system becomes strongly dependent on the electric dipole moment of the biomolecule by 
means of both the dipolar and Born attractive interactions.
Hence, for small values of $\mub$, the total PMF is repulsive and no molecule adsorption takes place in the system, with the exception of extreme values of $\qb$. In the case of very low molecule net charge, monopolar and osmotic repulsions are not strong enough to prevent stable surface adsorption due to the dipolar attraction and, to a lesser extent, the Born solvation energy contribution. On the other hand, for the highest values of $\qb$ the system exhibits metastable adsorption as a result of the competition between the repulsive monopolar interaction and the attractive Born contribution, given that the electrostatic dipolar interaction does not depend on the biomolecule net charge. Indeed, at high values of $\qb$ the difference in charging energy dramatically increases and can overcome the effect of monopolar repulsion at the hydrogel surface. As the biomolecule electric dipole moment increases, the repulsion due to the electrostatic monopolar interaction and the osmotic pressure remains constant for a 
given $\qb$. Therefore, the increase of the dipolar interaction with $\mub$ at the hydrogel surface and the Born contribution inside the hydrogel network lessen the total PMF, leading first to metastable surface adsorption of the biomolecule, then to stable adsorption, and finally to the partial sorption of the particle inside the hydrogel for the highest values of dipole moment. These trends with the net charge and the dipole moment of the biomolecule have also been observed in the encapsulation of charged proteins such as cytochrome-c, chymotrypsin and ribonuclease within oppositely-charged reverse micelles\cite{Bratko1988,Pitre1993,Chen2000,Pinero2004}.
Indeed, increasing the protein net charge enhances the encapsulation of the protein, whereas raising the dipole moment strongly enhances its binding to the micelle interface. Moreover, the addition of screening salt reduces the electrostatic binding of the protein, in a similar fashion to our theoretical predictions with porous hydrogels.

% Phase diagrams BVnonelec = 0

\begin{figure}[ht!]
 \centering
 \includegraphics[width=3.25in]{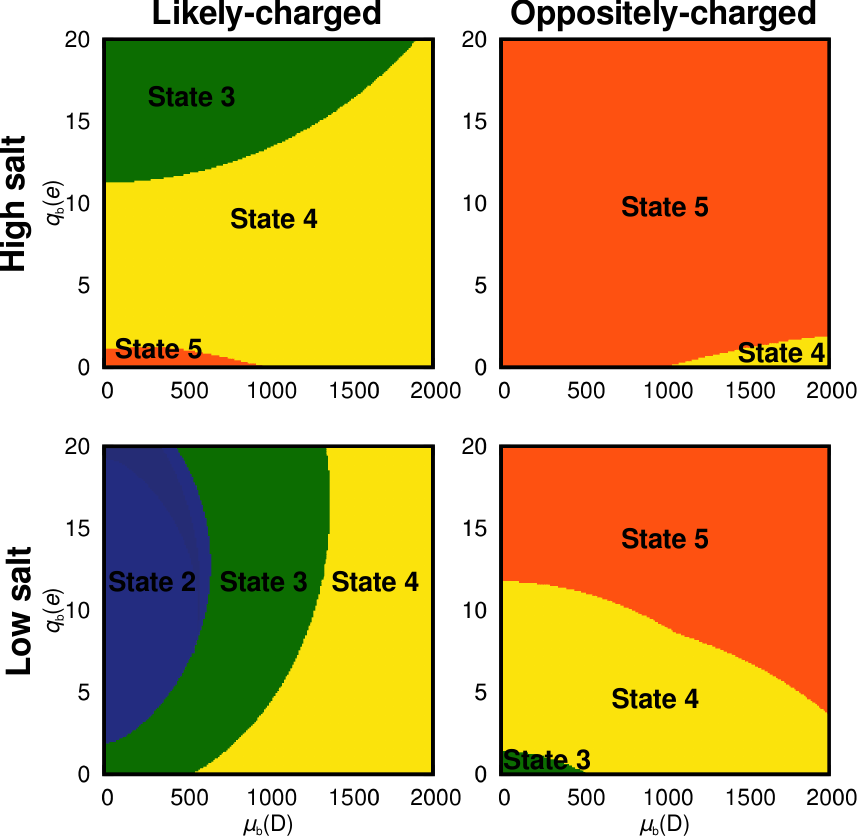}
 \caption{($\mub, \qb$) state diagrams obtained in the high and low salt concentration regimes (up and down panels, respectively) and for like-charged (left panels) and oppositely-charged (right panels) hydrogel and biomolecule in a system in which biomolecule hydrophobicity dominates the specific interaction ($\Vspec(r)<0$).}
 \label{fig:PhaseDiagramBVelec-5}
\end{figure}

Let us now focus on the influence of specific hydrophobic-steric effects. In Figure~\ref{fig:PhaseDiagramBVelec-5} we show the state diagrams for likely- and oppositely-charged biomolecules at high and low salt concentrations for a system in which the hydrophobic attraction dominates the specific interaction. In particular, we fixed the specific attraction at the hydrogel center ($r=0$) to be $\Vspeco=-5 \kB T$. The state diagram is now explored taking into account this new specific attraction together with the other energetic contributions discussed above. Therefore, it is expected an enhancement of the biomolecule sorption with respect to the system described in Figure~\ref{fig:PhaseDiagramBVelec0}. In fact, we can clearly see that there is always some kind of sorption in the four state diagrams of the hydrophobic system, while the state 1 of no-adsorption never takes place. In particular, for oppositely-charged particles, the biomolecule is always sorbed due to the strong attraction induced by the 
hydrogel. Only for very small values of net charge and dipole moment and at low salt concentrations, the osmotic pressure is repulsive enough to avoid the permeation inside the polyelectrolyte network, but leading instead to a stable adsorption of the biomolecule onto the hydrogel surface.

On the other hand, when the electric charge of the hydrogel and the biomolecule share the same sign, it has been mentioned that the sorption state at high electrolyte concentration is resulting from the interplay between the electrostatic monopolar repulsion and the dipolar attraction. However, in the case of hydrophobic molecules, the monopolar repulsion is overcome by this effective attraction due to the tendency of the particles to exclude water molecules. Hence, the resulting state diagram is highly dominated by the sorption state, although when the biomolecule net charge is high enough, the electrostatic repulsion restricts its sorption to the hydrogel surface, leading to state 3. Similar changes affect the state diagram of likely-charged particles in the low salt concentration regime: the repulsion due to the electrostatic monopolar interaction and the osmotic pressure is not enough to prevent the sorption of the hydrophobic biomolecule. Hence, even for small values of the electric dipole moment, the 
molecule experiences a metastable surface adsorption. As $\mub$ increases, the attraction due to the dipolar interaction at the hydrogel surface and the difference in Born solvation energy increases, leading to stable surface adsorption at mid-values of dipole moment and to full inside sorption of the particle for higher values of $\mub$.

% Phase diagrams BVnonelec = -5

\begin{figure}[ht!]
 \centering
 \includegraphics[width=3.25in]{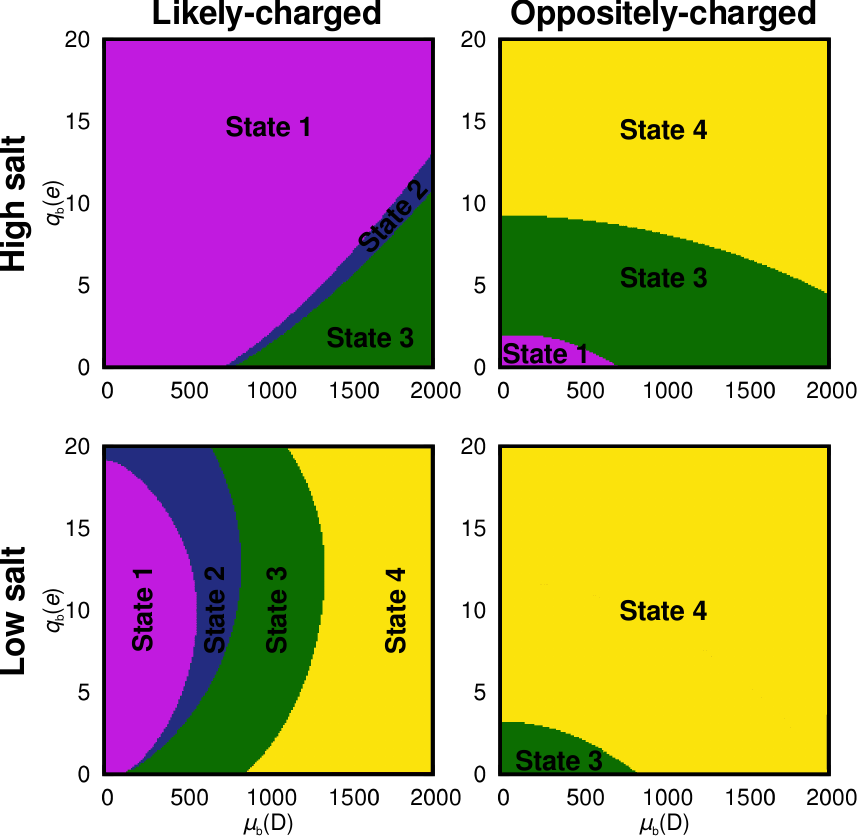}
 \caption{($\mub, \qb$) state diagrams obtained in the high and low salt concentration regimes (up and down panels, respectively) and for like-charged (left panels) and oppositely-charged (right panels) hydrogel and biomolecule in a system with strong excluded-volume effects ($\Vspec(r)>0$).}
 \label{fig:PhaseDiagramBVelec5}
\end{figure}

Let us finally consider the least favorable case for the biomolecule sorption, that is, a hydrogel with high excluded-volume effects. To represent such a system we have considered a repulsive specific potential featured by a repulsion of $\Vspeco=5 \kB T$ at the hydrogel center. The corresponding state diagrams are plotted in Figure~\ref{fig:PhaseDiagramBVelec5}, where we can easily observe that the steric exclusion leads to the opposite trend to the one predicted in the presence of hydrophobic attraction (described in Figure~\ref{fig:PhaseDiagramBVelec-5}). That is, while the inside sorption states were dominant in most cases in the presence of hydrophobic attraction, in this situation it looses ground in favour of other hydrogel-biomolecule configurations. Actually, at first sight the four state diagrams are quite similar to those of Figure~\ref{fig:PhaseDiagramBVelec0}, with subtle differences. Regarding the oppositely-charged systems, we can see that higher values of net charge are needed to attain the 
biomolecule sorption inside the polyelectrolyte network. This effect is more remarkable at high electrolyte concentration, where the sorption state of the biomolecule is the result of the competition between electrostatic attraction and steric repulsion. Moreover, for small enough values of net charge and dipole moment, the excluded-volume effects dominate the total hydrogel-biomolecule  interaction, avoiding any permeation of the molecule (state 1).

The same tendency is observed for likely-charged biomolecules, especially in the high electrolyte concentration regime, where the total PMF is mainly repulsive due to the joint effect of electrostatic monopolar interaction and steric exclusion.
Only for small values of the net charge and high values of the dipole moment, the electrostatic dipolar attraction becomes strong enough to allow meta- and stable surface adsorptions of the biomolecule.
Conversely, in the low salt concentration regime, the state of the system is principally controlled by the osmotic and Born energetic terms, so the volume exclusion represents an small perturbation, leading to a similar behavior than the one predicted for  $\Vspeco=0$.
This is a reasonable result because, in contrast to the high salt regime,the total PMF has a magnitude of tens of $\kB T$ at low salt concentration.
Consequently, a steric repulsion of $\Vspeco=5\kB T$ has little effect on the total interaction potential.

% Phase diagrams BVnonelec = 5

%-------------------------------------------------------------------------------------------------%
% Conclusions
%-------------------------------------------------------------------------------------------------%
\section{Conclusions}

The aim of this work was to understand the different mechanisms that are involved in the sorption and spatial distribution of a multipolar biomolecule to  a hydrogel particle in presence of a 1:1 electrolyte. To construct the total potential of mean force (PMF) between the hydrogel and the dipolar biomolecule, different phenomenological contributions were considered. These are the electrostatic interaction and Born solvation energies up to the dipolar term of the multipolar expansion, the osmotic pressure contribution, and a specific (excluded-volume or hydrophobic) interaction. From the study of the PMF, five sorption states of the system were specified, namely no adsorption, metastable surface adsorption, stable surface adsorption and partial and full inside sorption. The resulting sorption states were compiled in state diagrams as a function of net charge and electric dipole moment, for different electrolyte concentrations and specific hydrogel-biomolecule interactions.

Results show that for oppositely-charged particles, the biomolecule was sorbed inside the polyelectrolyte network for almost every combination of net charge and dipole moment at both high and low salt concentration limits, with the exception of very small values of ($\mub, \qb$) and high excluded-volume effects. In those less favorable situations the biomolecule experienced a stable surface adsorption or, just in case of really high steric repulsion, it was not adsorbed.

For likely-charged hydrogel-biomolecule, however, the electrolyte concentration had a determinant effect on the overall configuration of the state diagrams, which showed more variability. At high electrolyte concentration, the sorption state of the biomolecule was mainly the result of the interplay between the electrostatic monopolar repulsion and the dipolar attraction. Consequently, in most cases the biomolecule is precluded by the hydrogel network, except for hydrophobic particles. In the rest of cases, high values of the electric dipole moment were needed to achieve some kind of sorption of the molecule. On the other hand, in the low salt regime the adsorption of the biomolecule was subject to a multiple competing interaction mechanisms, being the electric dipole moment the driving variable that leads the system to all the different sorption states, from complete repulsion to full inside sorption.

From this study we can conclude that the electrostatic interaction has a determinant role in the uptake and spatial distribution of charged molecules inside hydrogels, as has been suggested by several authors~\cite{Bysell2009,Bysell2009b,Hansson2012,Smith2011}. Although biomolecules of the opposite charge of the hydrogel are more feasible to be adsorbed, we have seen that different sorption states can be achieved even for likely-charged biomolecules. 
{\it Importantly, the non-uniform charge distribution of the biomolecule leading to large dipoles is is a determining factor to take into account, especially when surface adsorption is a focus.} We emphasize again that the protein charge distribution is tuneable by pH. Further, the hydrophobicity of the particle clearly enhances the biomolecule uptake inside the polyelectrolyte network.

In summary, there is no a unique variable which controls the sorption of molecules inside the hydrogel, but a wide range of factors are to consider. The total hydrogel-biomolecule interaction potential is the result of a complex combination of mechanisms that depend on the molecule net charge and electric dipole moment, electrolyte bulk concentration, hydrogel charge, monomer distribution, volume, and hydrophobic character of the particles, among others. Although the complexity of the problem may seem discouraging, we have demonstrated that it is possible to predict the qualitative behavior of such systems with the help of a relatively simple theoretical framework. Not only discerning whether the molecule is taken up by the hydrogel or not, but determining the place where the particle is located within the carrier particle is crucial for both the encapsulation and release kinetics of the biomolecule and hence hydrogel carrier functionality. Therefore, our work should serve as a useful guide in the 
development of hydrogel-based carrier systems for biotechnological applications due to its simplicity and effectiveness.

In future extensions of our theory we intend to further investigate the interplay between internal absorption and surface adsorption on the equilibrium distribution of many (interacting) biomolecules as well as on the kinetic properties of the sorption process, making special emphasis on the effect of the finite biomolecule concentration. In this regard, experimental results show that the accumulation of biomolecules at the external shell of the hydrogel can behave as a steric barrier for further permeation of cosolute into the hydrogel core~\cite{Bysell2006}. Therefore, it would be interesting to extend our model to these more complex situations for different hydrogel charge distributions, namely, uniformly charged or surface charged hydrogels.

\begin{acknowledgement}

The authors thank the computational resources provided by \textsc{Proteus} from the Institute Carlos I for Theoretical and Computational Physics (University of Granada) and by the Institute for Soft Matter and Functional Materials (Helmholtz-Zentrum Berlin). I.A.-B. and A.M.-J. acknowledge funding by the Spanish \textquoteleft Ministerio de Econom\'{\i}a y Competitividad (MINECO), Plan Nacional de Investigaci\'{o}n, Desarrollo e Innovaci\'{o}n Tecnol\'{o}gica (I+D+i)' (Project FIS2016-80087-C2-1-P). J.D. thanks Matthias Ballauff and Stefano Angioletti-Uberti for inspiring discussions and acknowledges support by the the ERC (European Research Council) Consolidator Grant under project 646659-NANOREACTOR. 

\end{acknowledgement}

%-------------------------------------------------------------------------------------------------%
\section{Appendix: Electrostatic potential and electric field}
\label{sec:Appendix1}
%-------------------------------------------------------------------------------------------------%

The full Poisson-Boltzmann (PB) equation in spherical coordinates is given by
\begin{equation}
 \resizebox{0.86\linewidth}{!}{$\bigtriangleup \psi(r) = \displaystyle\frac{\d^2 \psi}{\d r^2} + \displaystyle\frac{2}{r}\frac{\d \psi}{\d r} = \begin{cases}
   \displaystyle-\frac{\rs(r)+\rh(r)}{\epsr' \epso}, & 0 < r < \Rh\\
   \displaystyle-\frac{\rs(r)}{\epsr \epso}, & \Rh < r < +\infty
 \end{cases},$}
 \label{}
\end{equation}
where $\rho_s = ec_s = e(c_+-c_-)$ and $\rho_h = e c_h$ are the charge densities of the salt and gel (with monovalent monomers), 
respectively. We assume the same relative permittivity of water inside and outside the swollen hydrogel particle ($\epsr' = \epsr = 80$).
Since the electrolyte ions obey Boltzmann's law,
\begin{equation}
 \rs(r) = \sum_{i}^{N_{\rm ions}} \zi e c_i^{\rm bulk} \exp\left( -\frac{\zi e \psi(r)}{\kB T} \right),
 \label{eq:BoltzmannLaw}
\end{equation}
the charge density at $r$ in the case of a symmetric 1:1 electrolyte with valence $z_\pm = \pm 1$ and bulk concentration $\csbulk$ is given by
\begin{equation}
 \rs(r)= -2  e \csbulk \sinh[\b  e \psi(r)]. 
% = z e \csbulk \left(\e^{-\b\zi e \psi(r)} - \e^{\b\zi e \psi(r)} \right) 
 \label{eq:rel_1-1}
\end{equation}
Therefore, we obtain the following Poisson-Boltzmann equation:
\begin{equation}
  \resizebox{0.86\linewidth}{!}{$\displaystyle\frac{\d^2 \psi}{\d r^2} + \displaystyle\frac{2}{r}\frac{\d \psi}{\d r} = \begin{cases}
   \displaystyle\frac{2 e \csbulk}{\epsr \epso} \sinh\left[\frac{e \psi(r)}{\kB T} \right] - \frac{e \ch(r)}{\epsr \epso}, & 0 < r < \Rh\\
   \\
   \displaystyle\frac{2 e \csbulk}{\epsr \epso} \sinh\left[\frac{e \psi(r)}{\kB T} \right], & \Rh < r < +\infty
 \end{cases}.$}
 \label{}
\end{equation}

We define the scaled electrostatic potential as $y(r) \equiv e\psi(r) / \kB T$.
In the low potential limit we can linearize the PB equation, so that: $|y(r)| \ll 1 \ \Rightarrow \ \sinh(y) \simeq y.$
This yields to the Debye-H\"{u}ckel equation:
\begin{equation}
 \resizebox{0.86\linewidth}{!}{$y''(r) + \displaystyle\frac{2}{r} y'(r) = \begin{cases}
   \displaystyle \k^{2} y(r) - \k^{2} \frac{\ch(r)}{2 \csbulk}, & 0 < r < \Rh\\
   \displaystyle \k^{2} y(r), & \Rh < r < +\infty
 \end{cases},$}
 \label{eq:linearizedPB}
\end{equation}
where $\k$ is the the inverse of the Debye screening length, also known as Debye-H\"{u}ckel parameter:
$$\k = \left( \frac{2 z^2 e^2 \csbulk}{\epsr \epso \kB T} \right)^{1/2}.$$

If the electrolyte bulk concentration is small, the Debye screening length $\lD$ is much greater than the intrinsic hydrogel-wather interface width, $\s$.
Hence, in the limit of low salt concentration we can assume a vanishing $\s$, that is, a step charge distribution inside the hydrogel particle:
\begin{equation}
\ch(r) = \begin{cases}
 \cho, & 0 < r < \Rh\\
 0, & \Rh < r < +\infty
\end{cases}.
\label{eq:cmrLowSalt}
\end{equation}

In this limit the linearized PB equation is given by
\begin{equation}
 \resizebox{0.86\linewidth}{!}{$y''(r) + \displaystyle \frac{2}{r} y'(r) = \begin{cases}
   \displaystyle \k^{2} y(r) - \k^{2} \yo, & 0 < r < \Rh\\
   \displaystyle \k^{2} y(r), & \Rh < r < +\infty
 \end{cases},$}
 \label{eq:PB_LowSalt}
\end{equation}
where $\yo \equiv \displaystyle\frac{\cho}{2 \csbulk}$ is the electrostatic potential at $r=0$.

On the other hand, in the limit of high salt concentration, the hydrogel-water interface width is comparable to Debye length.
In this case we can approximate the charge distribution as a error function, so that
\begin{equation}
 \resizebox{0.86\linewidth}{!}{$\ch(r) = \displaystyle \frac{\cho}{2} \left[ 1 - \epsrf\left( \frac{r-\Rh}{\sqrt{2} \ \s} \right) \right], \ 0 < r < +\infty.$}
 \label{eq:cmrHighSaltApp}
\end{equation}

In this case the linearized PB equation is given by
\begin{flalign}
\begin{split}
 & y''(r) + \frac{2}{r}y'(r) = \\
 & \resizebox{0.86\linewidth}{!}{$\displaystyle \k'^{2} y(r) - \k^{2} \yo\left[ 1 - \epsrf\left( \frac{r-\Rh}{\sqrt{2} \ \s} \right) \right],  0 < r < +\infty.$}
 \label{eq:PB_erf}
\end{split}
\end{flalign}

\subsection{Electrostatic potential}

In the low electrolyte concentration limit is possible to analytically solve the linearized Poisson-Boltzmann equation \eqref{eq:PB_LowSalt}.
This equation was solved by Ohshima\cite{Ohshima2008} for a spherical soft core-shell particle.
The solution for our model is a particular case of this previous work:
\begin{equation}
   \resizebox{0.86\linewidth}{!}{$y(r) = \begin{cases}
   \displaystyle \yo \left[ 1 - (\k\Rh + 1) \e^{-\k\Rh} \frac{\sinh(\k r)}{\k r} \right], & 0 < r < \Rh \\
   \\
   \displaystyle \yo\Rh [\cosh(\k\Rh) - \sinh(\k\Rh)] \frac{\e^{-\k r}}{\k r} , & \Rh < r < +\infty
 \end{cases}.$}
 \label{eq:yLowSalt}
\end{equation}

In the high electrolyte concentration limit, however, the solution is much more straightforward.
If the thickness of the charge layer is much greater than the Debye length ($\s \gg \lD$),
then the potential deep inside the particle becomes the Donnan potential $\pD$,
which is obtained by setting electroneutrality at $r=0$:
\begin{flalign}
\begin{split}
  & \pD = \frac{\kB T}{z e} \textrm{\ arcsinh} \left( \frac{Z \cho}{2 z \csbulk} \right) = \\
  & \frac{\kB T}{z e} \ln \left[ \frac{Z \cho}{2 z \csbulk} + \sqrt{1 + \left( \frac{Z \cho}{2 z \csbulk} \right)^{2}} \right]
  \label{eq:Donnan_full}
\end{split}
\end{flalign}

In the limit of large electrolyte concentration [$\cs(r) \gg \ch(r)$] we can linearize the Donnan potential,
\begin{equation}
 \pD \simeq \frac{\kB T}{e} \frac{\cho}{2 \csbulk}.
\end{equation}
In this regime the electrostatic local effects decay very fast with distance to the hydrogel center, so electroneutrality is fulfilled at any $r$:
\begin{equation}
 y(r) \simeq \frac{\ch(r)}{2 \csbulk} = \frac{\yo}{2} \left[ 1 - \epsrf\left( \frac{r-\Rh}{\sqrt{2} \ \s} \right) \right].
\end{equation}

\subsection{Electric field}

We define the scaled electric field as $Y(r) \equiv  e E(r) / \kB T$. 
Given that the system under study has spherical symmetry, the electric field can simply be obtained from the potential:
\begin{equation}
 \mathbf{E}(r) = - \nabla \psi(r) = \frac{\d \psi(r)}{\d r} \hat{r}.
\end{equation}

Hence, in the low electrolyte concentration limit:
\begin{equation}
   \resizebox{0.86\linewidth}{!}{$Y(r) = \begin{cases}
   \displaystyle \yo(\k\Rh + 1)\e^{-\k\Rh} \displaystyle\frac{\k r \cosh(\k r) - \sinh(\k r)}{\k r^2}, & 0 < r < \Rh \\
   \\
   \displaystyle \yo [\k\Rh \cosh(\k\Rh) - \sinh(\k\Rh)]\displaystyle\frac{(\k r + 1) \e^{-\k r}}{\k r^2}, & \Rh < r < +\infty
 \end{cases}.$}
 \label{eq:YrLowSalt}
\end{equation}

In the high electrolyte concentration limit:
\begin{equation}
 Y(r) = \frac{\yo}{\sigma \sqrt{2\pi}} \e^{-\displaystyle\frac{(r-\Rh)^2}{2\s^2}}.% \ \hat{r}
 \label{eq:YrHighSalt}
\end{equation}

\providecommand{\latin}[1]{#1}
\providecommand*\mcitethebibliography{\thebibliography}
\csname @ifundefined\endcsname{endmcitethebibliography}
  {\let\endmcitethebibliography\endthebibliography}{}

\end{document}